\documentclass[twocolumn,showpacs,aps]{revtex4}
\usepackage[british]{babel}
\usepackage{graphicx}
\usepackage{amsmath}

\begin{document}

\title{Numerical simulations on the motion of atoms travelling
through\\a standing-wave light field}

\author{S.J.H.~Petra}
\email{stefan@nat.vu.nl}
\author{K.A.H.~van~Leeuwen}
\author{L.~Feenstra}
\thanks{\emph{Present address}: Physikalisches Institut, Universit\"at Hei\-del\-berg, Philosophenweg 12, 69120 Heidelberg, Germany.}
\author{W.~Hogervorst}
\author{W.~Vassen}

\affiliation{Atomic and Laser Physics Group, Laser Centre Vrije Universiteit, De Boelelaan 1081, 1081 HV Amsterdam, The Netherlands}

\begin{abstract}
The motion of metastable helium atoms travelling
through a standing light wave is investigated with a
semi-classical numerical model. The results of a calculation
including the velocity dependence of the dipole force are compared
with those of the commonly used approach, which assumes a
conservative dipole force. The comparison is made for two atom
guiding regimes that can be used for the production of
nanostructure arrays; a low power regime, where the atoms are
focused in a standing wave by the dipole force, and a higher power
regime, in which the atoms channel along the potential minima of
the light field. In the low power regime the differences between
the two models are negligible and both models show that, for
lithography purposes, pattern widths of 150~nm can be achieved. In
the high power channelling regime the conservative force model,
predicting 100~nm features, is shown to break down. The model that
incorporates velocity dependence, resulting in a structure size of
40~nm, remains valid, as demonstrated by a comparison with quantum
Monte-Carlo wavefunction calculations.
\end{abstract}

\pacs{02.60.Cb, 32.80.Lg, 81.16.Rf}

\maketitle

\section{Introduction}
\label{intro}

The dipole force in a standing-wave light field has been used to
create nanoscale patterns with beams of neutral atoms for many
years. The standing wave can act as an array of optical lenses to
focus the atoms during deposition onto a substrate, thereby
creating a one-dimensional structure. The first experiments using
this technique were performed ten years ago with sodium
\cite{timp92} and chromium \cite{mccl93} atoms. Since then, atom
lithography with aluminium \cite{mcgo95}, cesium \cite{liso97},
metastable argon \cite{john98}, and metastable neon \cite{enge99}
has also been reported. For an overview of atom lithography,
see Ref.~\cite{mesc03}. The results of a number of these
experiments have been compared with semi-classical numerical
calculations \cite{berg94,mccl95}. These calculations are based on
a dipole force that is derived from an optical potential to focus
the atoms in the standing light wave. This force is conservative
and does not take into account the velocity dependence.
Quantum-mechanical calculations have been performed as
well~\cite{lee00}. These calculations do not suffer from this
restriction. However, they are computationally intensive and
describe the atomic motion in one dimension only.

The purpose of this paper is to investigate in detail the 3D
atomic motion through a standing-wave light field using a
semi-classical approach with a non-conservative, velocity
dependent dipole force. The results of these simulations are
compared with calculations using the conventional model with a
conservative dipole force. The model presented in this paper
follows the approach of Minogin and Serimaa \cite{mino79}, where
the dipole force is represented in the form of a Fourier series,
with coefficients that are calculated with a continued fraction
method. Previously, a similar model was used to calculate a
one-dimensional beam profile of sodium atoms in the far field
\cite{li94,li96}. The model that is presented here describes a
full 3D simulation of the motion of metastable helium atoms
through a standing light wave for the purpose of atom lithography
applications. Not only the usual low power regime, where the atoms
are focused in the standing light wave, is investigated, but also
a higher power regime. In the latter regime the atoms channel
through the standing wave, undergoing a damped oscillation around
the potential minima of the light field. This is of special
interest for lithography, as it allows the production of
nanostructures with relative insensitivity to the exact alignment
of atomic beam, substrate, and light field. It is, however, also at
these higher laser powers that the potential model is expected to
break down and the inclusion of the velocity dependence of the
force is essential. Experiments in this novel regime with
metastable helium atoms have been performed and, in a future
paper, the experimental pattern widths will be compared with the
calculations presented in this paper. With metastable helium
atoms, nanoscale structures can be created in a gold film on a
silicon substrate via a two-step process \cite{nowa96}. First, the
high internal energy of the focused helium atoms (20~eV) in the
$2~^3\textrm{S}_1$ metastable state is used to selectively damage
an organic resist layer through the standing light wave. Next, the
pattern is transferred to the underlying gold film by means of a
wet etching process.

In the simulations, the light field is blue detuned from the
$2~^3\textrm{S}_1 \rightarrow 2~^3\textrm{P}_2$ optical transition
($\lambda = 1083$~nm) of the helium atom. The atoms are therefore
attracted to the intensity minima of the standing light wave,
which minimizes spontaneous emission of photons by the atoms. The
atomic motion is calculated using only the dipole force. The
atomic momentum diffusion due to fluctuations of the dipole force
is neglected. The momentum diffusion is caused by the variation in
the number of absorbed and emitted photons, and by the random
direction of the spontaneously emitted photons. Furthermore, the
atom is assumed to remain in a steady state, i.e., transient
effects of the dipole force are neglected. In order to investigate
the effects of momentum diffusion and transient effects, some
results from Monte Carlo Wave Function (MCWF) calculations similar
to Lee \cite{lee00} are presented as well. These calculations
include the velocity dependence of the force, momentum diffusion,
atomic diffraction as well as transient effects. However, they are
one-dimensional and require significant computational resources.

In the next section, a general expression for the dipole force is
derived, and both semi-classical models are outlined.
Sec.~\ref{numsim} describes the method and the parameters of the
numerical simulation. The results of the semi-classical
simulations are presented in Sec.~\ref{results}, and they are
compared with the MCWF simulations. Finally, concluding remarks
are given in Sec.~\ref{concl}.

\section{Theoretical models}
\label{theomod}

\subsection{Dipole force of a standing wave}

The light force experienced by a two-level atom due to the
presence of a classically described light field can be determined
by calculating the change in momentum $\vec{p}$ of the atom under
influence of the Hamiltonian:
\begin{equation}
H = H_A - \vec{d}\cdot\vec{E},
\end{equation}
where $H_A$ contains the internal and kinetic energy of the atom
and $\vec{d}\cdot\vec{E}$ is the dipole interaction operator that
describes the coupling between the atomic dipole moment $\vec{d}$
and the electric field component $\vec{E}$ of the light field.
According to the Ehrenfest theorem, the quan\-tum-me\-chan\-i\-cal
analogue of the radiation force is given by the Heisenberg
equation of motion
\begin{equation}
\label{qmforce} \left<\vec{\mathcal{F}}\right> = \frac{d
\left<\vec{p}\right>}{d\,t} =
\frac{i}{\hbar}\left<[H,\vec{p}]\right> =
\left<\mbox{\boldmath$\nabla$}(\vec{d}\cdot\vec{E})\right> =
\left<\vec{d}\right>\cdot\mbox{\boldmath$\nabla$}\vec{E}.
\end{equation}
In the last step of Eq.~(\ref{qmforce}) the expectation value of
the electric field operator $\vec{E}$ is replaced by the value at
the atomic centre of mass. This is legitimate in the electric
dipole approximation, where the wavelength of the light field
$\lambda$ is large compared to the de Broglie wavelength
$\lambda_{dB} = \hbar/|\vec{p}|$ of the atom, and spatial
variations of the electric field on the scale of the atomic
wave-packet can be neglected.

The expectation value of the electric dipole operator $\vec{d}$
can be written in terms of the atomic density matrix $\rho$, which
describes the quantum-mechanical state of the two-level atom, as
\begin{eqnarray}
\langle\vec{d}\rangle & = & \mbox{Tr}(\rho\,\vec{d}) = \vec{d}_{ge} (\rho_{ge} + \rho_{eg})
\nonumber\\
& = & 2\vec{d}_{ge}(u(t) \cos{\omega t} - v(t) \sin{\omega t}),
\end{eqnarray}
where the atomic density matrix elements $\rho_{ge}$ and
$\rho_{eg} = \rho_{ge}^*$ are the electronic coherences between
the ground state and the excited state of the atom, $\omega$ is
the frequency of the radiation field, and $u(t)$ and $v(t)$ are
two components of the Bloch vector. In the rotating-wave
approximation, where non-resonant terms of the atom-light
interactions are neglected, the components of the Bloch vector can
be written as
\begin{eqnarray}
u(t) & = & \frac{1}{2}(\rho_{ge}e^{i \omega t} + \rho_{eg}e^{-i \omega t}),
\nonumber\\
v(t) & = & \frac{1}{2i}(\rho_{ge}e^{i \omega t} - \rho_{eg}e^{-i \omega t}),
\nonumber\\
w(t) & = & \frac{1}{2}(\rho_{ee} - \rho_{gg}),
\end{eqnarray}
where the atomic density matrix elements $\rho_{gg}$ and
$\rho_{ee}$ are the populations of the ground state and the
excited state of the atom, normalized to $\rho_{gg} + \rho_{ee} =
1$. For a travelling wave, the electric field component of the
light field is given by
\begin{equation}
\vec{E}(\vec{r},t) = \mbox{\boldmath$\vec{\varepsilon}$}(\vec{r})\,E_0(\vec{r})\cos{(\omega t + \Phi(\vec{r}))},
\end{equation}
where \boldmath$\vec{\varepsilon}$\unboldmath$(\vec{r})$,
$E_0(\vec{r})$, and $\Phi(\vec{r})$ $(= -\vec{k}\cdot\vec{r}$) are
the polarization, amplitude and phase of the light wave
respectively at the atomic centre-of-mass position $\vec{r} =
(x,y,z)$. The Rabi frequency $\Omega(\vec{r})$ of the light field
is then defined as
\begin{equation}
\Omega(\vec{r}) =
-\frac{\vec{d}_{ge}\cdot\mbox{\boldmath$\vec{\varepsilon}$}(\vec{r})\,E_0(\vec{r})}{\hbar}.
\end{equation}
The general expression for the light force (in the electric dipole
and rotating-wave approximations) can be written as \cite{cook79}
\begin{equation}
\label{generalforce} \left<\vec{\mathcal{F}}\right> =
\vec{F}(\vec{r}) = -\hbar
u_{st}\mbox{\boldmath$\nabla$}\Omega(\vec{r}) -
\hbar\Omega(\vec{r})v_{st}\mbox{\boldmath$\nabla$}\Phi(\vec{r}).
\end{equation}
The two parts on the right-hand side of Eq.~(\ref{generalforce})
are the dipole force, proportional to the gradient of the Rabi
frequency $\Omega(\vec{r})$, and the scattering force,
proportional to the gradient of the phase $\Phi(\vec{r})$ of the
light field. The Bloch vector components $u(t)$ and $v(t)$ are
replaced by their time-independent steady-state values $u_{st}$
and $v_{st}$ respectively. This is valid in the adiabatic
approximation, where the atom moves slowly enough in the light
field to maintain an equilibrium between its internal state and
the radiation field.

The steady-state values can be found by solving the equations of
motion of the optical Bloch vector that describe the time
evolution of a two-level atom in a light field
\begin{equation}
\label{obe}
\left(\begin{array}{c}\dot{u}\\ \dot{v}\\ \dot{w} \end{array} \right)
=
\left(
\begin{array}{ccc}
-\Gamma/2 & \Delta + \dot{\Phi} & 0 \\
-(\Delta + \dot{\Phi}) & -\Gamma/2 & -\Omega \\
0 & \Omega & -\Gamma
\end{array}
\right)
\left(\begin{array}{c}u\\ v\\ w \end{array} \right)
-
\left(\begin{array}{c}0\\ 0\\ \Gamma/2 \end{array}
\right),
\end{equation}
where $\Gamma/2\pi$ is the natural linewidth of the excited state
and $\Delta/2\pi$ is the detuning between the light field
frequency $\omega$ and the frequency of the atomic transition.

The electric field component of a standing-wave light field,
composed of an incident and back-reflected Gaussian wave
travelling in the $x$-direction, can be written as
\begin{equation}
\vec{E}(\vec{r},t) = \mbox{\boldmath$\vec{\varepsilon}$}\,E_0
\cos{(\omega t)} \sin{(k x)}
\exp\left(-\frac{y^2}{w_{y}^2}-\frac{z^2}{w_{z}^2}\right),
\end{equation}
where $k$ is the wave number of the light and $w_y$ and $w_z$ are
the waists of the Gaussian beam profile in the $y$ and $z$
direction respectively. Since this electric field has no phase
dependence, the phase gradient term in Eq.~(\ref{generalforce})
vanishes. The general solution of the light force in a
standing-wave light field then becomes
\begin{equation}
\vec{F}_{sw}(\vec{r}) = -\hbar u_{st}\mbox{\boldmath$\nabla$}\Omega(\vec{r}).
\label{sw}
\end{equation}
The steady-state value of the optical Bloch vector component
$u_{st}$ can be derived using two different approaches; one that
neglects the atomic velocity (Sec.~\ref{potmodel}), and one that
is valid for arbitrary atomic velocities (Sec.~\ref{cfmodel}).

\subsection{Atom at rest -- potential model}
\label{potmodel}

When the transverse velocity of the atom in a standing-wave light
field is negligible ($\vec{k}\cdot\vec{v} \ll \Gamma$), the atom
travels over a very small distance compared to the optical
wavelength $\lambda$ during the relaxation time $\Gamma^{-1}$ of
the atom. The optical Bloch equations given in Eq.~(\ref{obe}) can
then be considered as a set of coupled linear differential
equations with time-independent coefficients. The steady-state
solutions are found analytically by setting $\dot{u} = \dot{v} =
\dot{w} = 0$ and they are given by
\begin{eqnarray}
\label{stsolution}
u_{st}(\vec{r}) & = & \frac{\Delta}{\Omega(\vec{r})}\frac{s(\vec{r})}{1+s(\vec{r})},
\nonumber\\
v_{st}(\vec{r}) & = & \frac{\Gamma}{2 \Omega(\vec{r})}\frac{s(\vec{r})}{1+s(\vec{r})},
\nonumber\\
w_{st}(\vec{r}) & = & -\frac{1}{2(1+s(\vec{r}))},
\end{eqnarray}
where
\begin{equation}
s(\vec{r}) = \frac{2\Omega^2(\vec{r})}{\Gamma^2 + 4\Delta^2}
\end{equation}
is the saturation parameter. The final expression for the dipole
force acting on an atom at rest in a standing-wave light field,
can now be found by combining Eq.~(\ref{sw}) and
Eq.~(\ref{stsolution})
\begin{equation}
\vec{F}_{pot}(\vec{r}) = -\hbar\Delta
\frac{\mbox{\boldmath$\nabla$}
\Omega^2(\vec{r})}{2\Omega^2(\vec{r}) + \Gamma^2 + 4\Delta^2}.
\end{equation}
This force is conservative and it can be written as the gradient
of a potential \cite{cook79,ashk78}
\begin{equation}
\label{fpot} \vec{F}_{pot}(\vec{r}) = -\mbox{\boldmath$\nabla$}
U(\vec{r}) = -\mbox{\boldmath$\nabla$}\frac{\hbar
\Delta}{2}\ln{\left[1+s(\vec{r})\right]}.
\end{equation}
Eq.~(\ref{fpot}) is a well-known expression for the dipole force
and it is commonly used for semi-classical calculations of atomic
motion in a standing-wave light field \cite{timp92,mccl93,mccl95}.
Since this force is conservative, the kinetic energy of the atom
at any moment is determined by the local potential of the light
field.

\subsection{Moving atom -- Minogin model}
\label{cfmodel}

When an atom moves with velocity $\vec{v} \neq 0$ in the
standing-wave light field, the position of the atom $\vec{r}$
becomes explicitly time dependent as $\vec{r}(t) = \vec{v} t$.
However, the time dependence of the transverse coordinates $y$ and
$z$ can be neglected, since the wavelength of the light $\lambda$
is much smaller than the waist of the Gaussian beam profile. This
means that the optical Bloch vector components $u$, $v$, and $w$
change more rapidly along the axis of the light field
($x$-direction) than in the transverse directions $y$ and $z$.
Therefore, for finding the steady-state solution of the optical
Bloch vector components, only the time dependence of the
$x$-coordinate has to be taken into account. The Rabi frequency
$\Omega(\vec{r})$ can then be written as a periodic function of
time
\begin{eqnarray}
\label{Rabi_t}
\Omega(\vec{r},t) & = & \Omega_0(y,z) \sin (k x)
\nonumber\\
& = & \Omega_0(y,z) \sin (k v_x t),
\end{eqnarray}
where $\Omega_0(y,z)$ is the peak Rabi frequency at the anti-nodes
of the standing wave, and $v_x$ is the velocity of the atom
parallel to the axis of the light field. With the Rabi frequency
given by Eq.~(\ref{Rabi_t}), the coefficients of the coupled
linear differential equations in Eq.~(\ref{obe}) become time
dependent. Consequently, the optical Bloch equations can no longer
be solved analytically. However, a steady-state solution of $u$,
$v$ and $w$ can be found by expanding each of them in a Fourier
series
\begin{equation}
\label{fourierexpansion}
h(\vec{r}) = \sum_{n=-\infty}^{\infty} h_n(y,z)\,e^{i n k x},
\end{equation}
where the common notation $h$ is used for $u$, $v$, and $w$, and
the quantities $h_n$ satisfy the reality condition
$h_{-n}=h_{n}^*$. By substituting the Fourier expansions
Eq.~(\ref{fourierexpansion}) and the expression for the Rabi
frequency given by Eq.~(\ref{Rabi_t}) in the optical Bloch
equations Eq.~(\ref{obe}), a set of recursive algebraic equations
is obtained for the optical Bloch vector components \cite{mino79}
\begin{eqnarray}
\label{recur}
(\Gamma/2 + i n k v_x) u_n & = & \Delta v_n,
\nonumber\\
(\Gamma/2 + i n k v_x) v_n & = & - \Omega_0 \left(w_{n-1} + w_{n+1}\right) - \Delta u_n,
\nonumber\\
(\Gamma + i n k v_x) w_n & = & \Omega_0 \left(v_{n-1} + v_{n+1}\right) - \frac{\Gamma}{2} \, \delta_{n0}.
\end{eqnarray}
The dipole force is now also represented in the form of a Fourier
series, where the Fourier coefficients can be calculated from the
optical Bloch vector components. The expression for the dipole
force can be separated into components parallel and perpendicular
to the axis of the stand\-ing-wave light beam ($x$-direction) as
\begin{eqnarray}
\label{Fcfpar}
\vec{F}_{\|}(\vec{r}) & = & \vec{F}_{\|}^{0}(\vec{r}) \\
& + & \sum_{n=1}^{\infty}(-1)^n \left( \vec{F}_{\| u}^{2
n}(\vec{r}) \cos{2 n k x} + \vec{F}_{\| v}^{2 n}(\vec{r}) \sin{2 n
k x} \right), \nonumber
\end{eqnarray}
and
\begin{eqnarray}
\label{Fcfperp}
\vec{F}_{\perp}(\vec{r}) & = & \vec{F}_{\perp}^{0}(\vec{r}) \\
& + & \sum_{n=1}^{\infty}(-1)^n \Bigl( \vec{F}_{\perp u}^{2
n}(\vec{r}) \cos{2 n k x} + \vec{F}_{\perp v}^{2 n}(\vec{r})
\sin{2 n k x} \Bigr), \nonumber
\end{eqnarray}
where the coefficients of the Fourier series are given by
\begin{eqnarray}
\label{parcoef}
\vec{F}_{\|}^{0}(\vec{r}) & = & -2\hbar k \Omega_0 \,\mbox{Im}\,u_1,
\nonumber\\
\vec{F}_{\| u}^{2 n}(\vec{r}) & = & -2\hbar k \Omega_0 \,\mbox{Im}\,(u_{2n+1} - u_{2n-1}),
\nonumber\\
\vec{F}_{\| v}^{2 n}(\vec{r}) & = & -2\hbar k \Omega_0 \,\mbox{Re}\,(u_{2n+1} - u_{2n-1}),
\end{eqnarray}
and
\begin{eqnarray}
\label{perpcoef} \vec{F}_{\perp}^{0}(\vec{r}) & = & 4\hbar
\left(\frac{y}{w_{y}^2}+\frac{z}{w_{z}^2}\right) \Omega_0
\,\mbox{Re}\,u_1,
\nonumber\\
\vec{F}_{\perp u}^{2 n}(\vec{r}) & = & 4\hbar
\left(\frac{y}{w_{y}^2}+\frac{z}{w_{z}^2}\right) \Omega_0
\,\mbox{Re}\,(u_{2n+1} + u_{2n-1}),
\nonumber\\
\vec{F}_{\perp v}^{2 n}(\vec{r}) & = & -4\hbar
\left(\frac{y}{w_{y}^2}+\frac{z}{w_{z}^2}\right) \Omega_0
\,\mbox{Im}\,(u_{2n+1} + u_{2n-1}),\qquad
\end{eqnarray}
The coefficients $u_n$ are obtained from the solution of the
recursion relations (Eq.~(\ref{recur})) in the form of convergent
continued fractions. The dipole force given by Eq.~(\ref{Fcfpar})
and Eq.~(\ref{Fcfperp}) is dissipative, and the atomic energy is
therefore not conserved. The momentum change of the atom can be
attributed to an additional damping or heating force, dependent on
the intensity of the light field and on the sign of its detuning
from atomic resonance.

\section{Numerical simulations}
\label{numsim}

\subsection{Calculation method} 
\label{calcmethod} 

The dipole force derived from the two models described in
Sec.~\ref{potmodel} and Sec.~\ref{cfmodel} is used to calculate
the atomic movement through a standing-wave light field. By
straightforward numerical integration of the Newtonian equations
of motion, the change of the atomic velocity and position under
influence of the dipole force at the current position of the atom
in the light field is calculated. In this way a full 3D simulation
of the atom trajectories and velocities in the standing-wave light
field is performed. The atomic pattern formation is mapped with 2D
atomic distribution plots in the $xy$-plane and in histograms.

The calculation of the coefficients $u_n$, required for the
calculation of the dipole force in the Minogin model, is described
in detail by Minogin and Letokhov \cite{mino87} and it is
summarized for completeness in the appendix. The solutions for the
coefficients are found in the form of convergent continued
fractions. For the simulations, these continued fractions are
calculated with an accuracy better than $10^{-16}$, which requires
a maximum of 5000 terms. This precision is necessary to calculate
the coefficients $u_n$ with an accuracy better than 1~ppm. The
Fourier series that represent the expression for the dipole force
contain at most 2500 Fourier coefficients. A convergence analysis
has shown that the dipole force change is negligible at higher
accuracies of the coefficients $u_n$ and when more terms are added
to the Fourier series. Therefore the level of convergence is
assumed to be sufficient.

\begin{figure*}[t] 
\includegraphics[width=\textwidth,keepaspectratio]{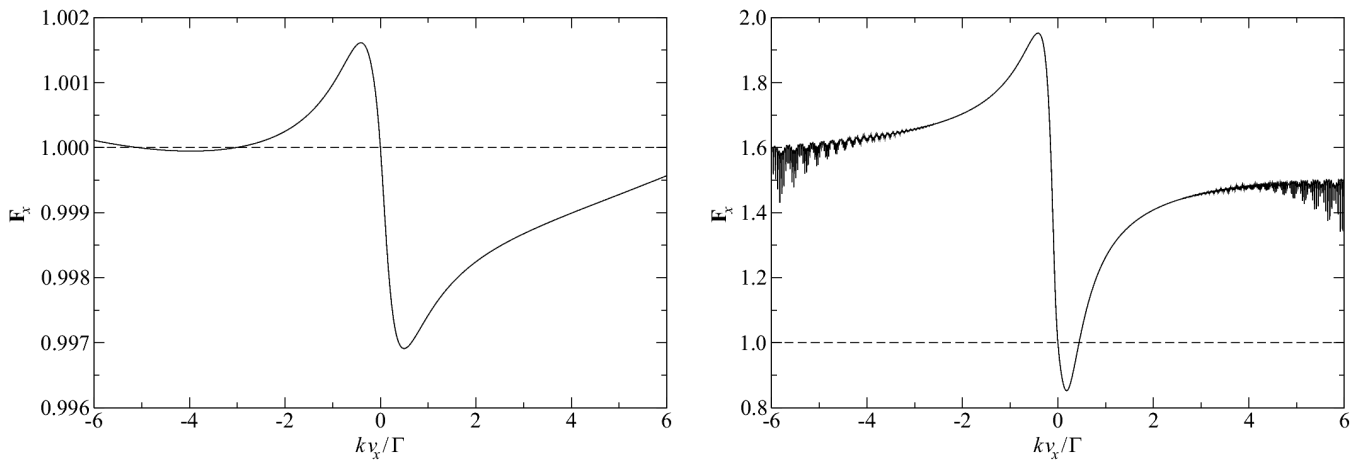} 
\caption {\label{velodep} Velocity dependence of the $x$-component 
of the dipole force of a standing-wave light field in the focusing 
regime (left) and the channelling regime (right) at position 
$(x,y,z)=(3\lambda/8,0,0)$ in the standing wave and at a detuning 
of $\Delta/2\pi = 375$~MHz. The dipole force according the Minogin 
model (solid lines) is normalized to the dipole force from the 
potential model (dashed lines).} 
\end{figure*}

\subsection{Parameters}

The calculations are performed on a beam of helium atoms in the
$2~^3\textrm{S}_1$ metastable state, which has a lifetime of about
8000~s. The atomic beam has a mean longitudinal velocity of
2000~m/s and a longitudinal velocity spread (full width at $1/e^2$
height) of 650~m/s. After collimation of the beam, the transverse
velocity spread of the atoms is reduced to about 3~m/s. For the
calculations, each atom is assigned a longitudinal and transverse
initial velocity that is randomly picked from Gaussian velocity
distributions with the above described averages and spreads.

The light of the standing wave has a wavelength of 1083~nm,
driving the $2~^3\textrm{S}_1 \rightarrow 2~^3\textrm{P}_2$
optical transition of the helium atom, which has a natural
linewidth $\Gamma/2\pi = 1.6$~MHz. By detuning the light field
relatively far to the blue side of the atomic resonance
($\Delta/2\pi = 375$~MHz), the atoms are attracted to the nodes of
the standing-wave light field, which reduces spontaneous
emissions. An upper limit for the detuning forms the
$2~^3\textrm{P}_1$ state, which energy level lies 2.3~GHz above
the $2~^3\textrm{P}_2$ state. For very large blue detunings of the
light field from the $2~^3\textrm{P}_2$ state, the atom can thus
interact with the light field via the $2~^3\textrm{S}_1
\rightarrow 2~^3\textrm{P}_1$ transition.

The Rabi frequency can be calculated from the intensity of the
light field as
\begin{equation}
\Omega(\vec{r}) = \Gamma\sqrt{\frac{I(\vec{r})}{2 I_{sat}}},
\end{equation}
where $I_{sat} = 0.17$~mW/cm$^2$ is the saturation intensity of
the optical transition and $I(\vec{r})$ is the intensity profile
of the standing-wave light field, given by
\begin{equation}
I(\vec{r}) = I_0 \sin^2{(k x)} \exp\left(-\frac{2 y^2}{w_{y}^2}-\frac{2 z^2}{w_{z}^2}\right).
\end{equation}
The Gaussian light beam has a circular beam profile with a waist
($1/e^2$ radius) $w_y = w_z = 331~\mu$m. The quantity $I_0$ is the
intensity of the light field at the anti-nodes of the standing
wave and it is given by
\begin{equation}
I_0 = \frac{8 P_0}{\pi w_y w_z},
\end{equation}
where $P_0$ is the power of the incident light beam. Depending on
this power, two different regimes can be distinguished for guiding
the atoms through the standing-wave light field. At low power, the
atoms can be focused at the centre of the Gaussian light beam. For
large detunings ($\Delta\gg\Gamma$), the power required for this
focusing can be calculated from \cite{mccl95}
\begin{equation}
\label{focpower}
P_0 = 5.37\frac{\pi m \vec{v}^2 \Delta I_{sat}}{2 \hbar \Gamma^2 k^2}.
\end{equation}
This focusing power is independent of the waist of the light beam.
For the conditions mentioned above, $P_0 = 2.4$~mW. For high-power
light fields, the atoms oscillate through a potential minimum of
the standing wave and the sign of the transverse velocity of the
atoms changes many times. This is called the channelling regime.
For calculations in this regime a power $P_0 = 800$~mW is used.
The dipole force in the channelling regime is then about one order
of magnitude larger than in the focusing regime.

\subsection{The dipole force}  

The velocity dependence of the $x$-component of the dipole force
in the focusing and channelling regime for both models is depicted
in Fig.~\ref{velodep}. The graphs show that, in the focusing
regime (left graph), the difference between the dipole force
according to the potential model and the Minogin model is very
small (at most 0.3\%). In the channelling regime (right graph),
however, the dipole force of the two models differs significantly.
The force shows a negative slope for small transverse velocities
($|k v_x/ \Gamma| < 0.2$), which means that the force is a damping
force. This cooling effect is essentially the ``blue-detuned
Sisyphus cooling'' introduced by Dalibard and Cohen-Tannoudji
\cite{dali85}. At larger velocities, the dipole force in the
Minogin model is significantly larger than in the potential model.
It is therefore expected that the distinction between the two
models will be most pronounced in the channelling regime.
Furthermore, in the channelling regime, the Minogin mod\-el shows
some resonance peaks at high atom velocities. These are called
Doppleron resonances \cite{kyro77}, and occur when the atom is
excited to the $2~^3\textrm{P}_2$ state by multiple photon
absorptions and emissions. When an atom absorbs ($n+1$) photons
from one wave of the standing-wave light field, and emits $n$
photons into the other, these resonances appear at velocities
\begin{equation}
k v_x/\Gamma = \pm\frac{\Delta}{(2n + 1)\Gamma}.
\end{equation}

\section{Results}
\label{results}

\begin{figure*}[t]
\includegraphics[width=\textwidth,keepaspectratio]{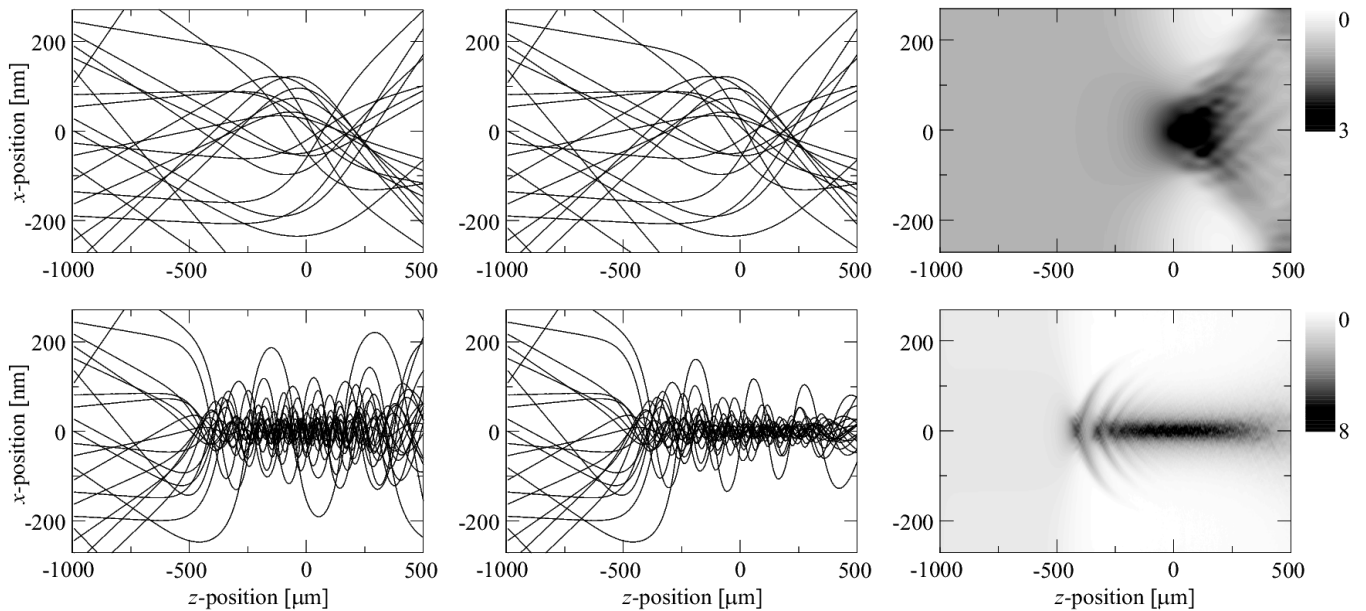}
\caption {\label{traject} Atom trajectories through the
standing-wave light field in the focusing regime (upper graphs)
and the channelling regime (lower graphs). Semi-classical
calculations are performed with the potential model (left) and the
Minogin model (centre). The right graphs show atomic density
distributions of Monte Carlo Wave Function (MCWF) simulations.}
\end{figure*}

Fig.~\ref{traject} shows a projection on the $xz$-plane of the
atomic motion through the standing-wave light field in the
focusing regime (upper graphs) and channelling regime (lower
graphs). The centre of the Gaussian light beam, which propagates
in the $x$-direction, is located at $z = 0$. The graphs cover half
a wavelength ($\lambda/2 = 542$~nm) of the standing-wave light
field, showing exactly one potential minimum through which the
atoms travel in the positive $z$-direction (from left to right).
To allow the atoms to experience a maximum dipole force, the
starting $y$-position is $y = 0$ for all atoms. However, since the
simulations are 3D and the atoms have an initial velocity in the
$y$-direction, this position is not maintained while travelling
through the light field.

Since the velocity dependence of the dipole force in the focusing
regime is almost negligible (see Fig.~\ref{velodep}), the atomic
motion in this regime according to the potential model (upper left
graph) is very similar to the calculation with the Minogin model
(upper centre graph). In the channelling regime, the calculated
trajectories differ considerably for the potential model (lower
left graph) and the Minogin model (lower centre graph), due to the
higher average value and the strong velocity dependence of the
dipole force in the latter model (see Fig.~\ref{velodep}). The
potential model is invalid in this regime, since the atoms have a
transverse velocity spread of 3~m/s ($k v_x/\Gamma = 1.7$), and
the transverse velocity of the atoms can even become on the order
of 10~m/s in the standing wave. In the channelling graph of
Fig.~\ref{traject}, the additional damping force results in an
increased confinement of the atoms in the potential well of the
standing wave. The upper right and lower right graphs of
Fig.~\ref{traject} show one-dimensional MCWF calculations for the
focusing regime and channelling regime respectively. In these
calculations atomic diffraction, velocity dependence and momentum
diffusion are fully incorporated. The plots show the atomic
density distribution on a grey-scale for 225 quantum trajectories,
corresponding to the same initial distribution as used for the
semi-classical simulations. The characteristic feather-like
structures in the channelling regime (lower right graph) do not
constitute a specific quantum feature. They correspond to
(approximate) caustics in the semi-classical trajectories and they
become clearly visible if a larger number of trajectories is
plotted in the same way as in the lower centre graph.

\begin{figure}[t]
\includegraphics[width=\columnwidth,keepaspectratio]{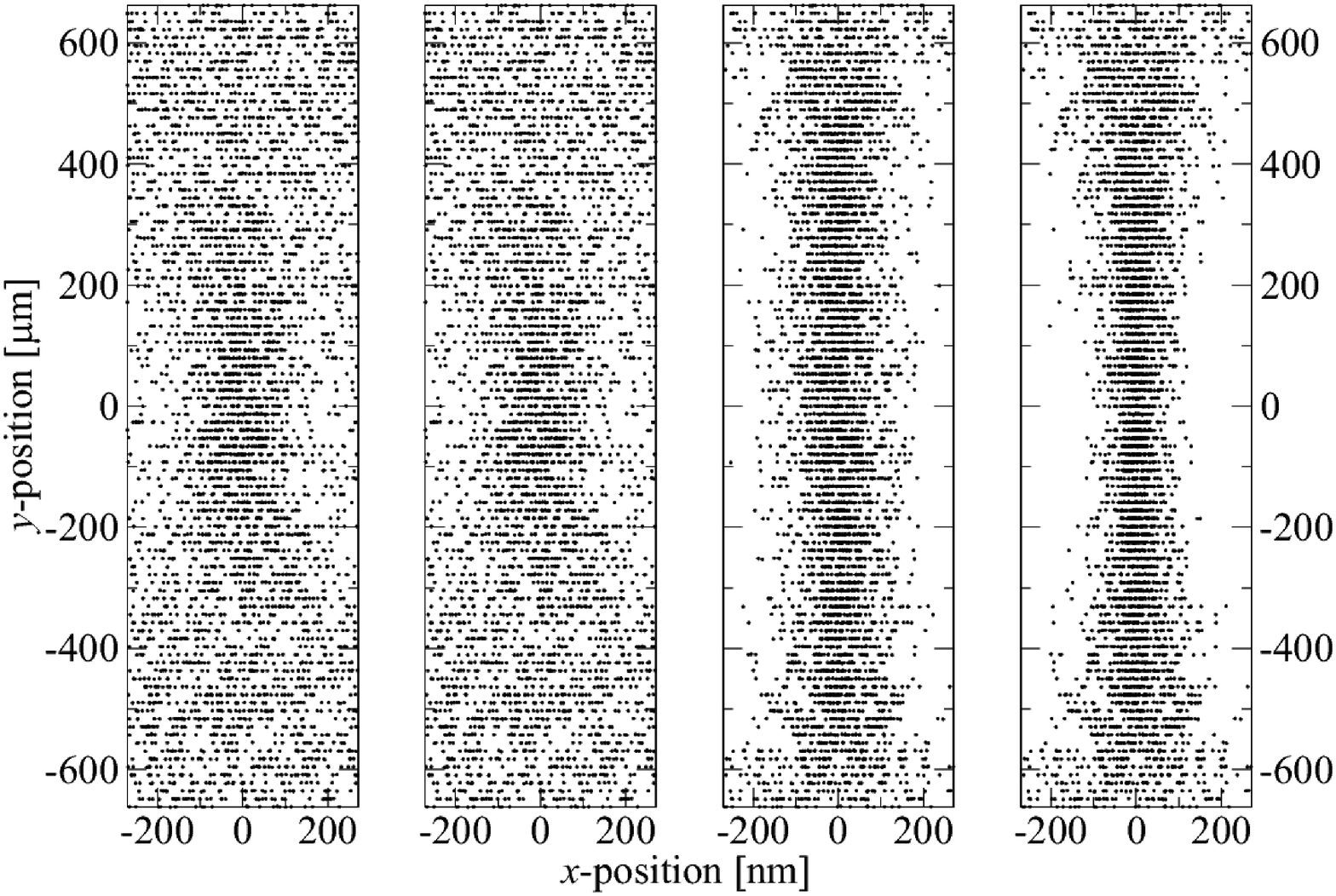}
\caption {\label{xyplots} Distribution plots of the positions
where the atoms hit the sample after travelling through the
standing-wave light field at sample position $z = w_z/4 =
83~\mu$m. From left to right, the first to graphs show plots of
the focusing regime using the potential model (first graph) and
the Minogin model (second graph). The last two graphs show the
channelling regime using the potential model (third graph) and the
Minogin model (fourth graph).}
\end{figure}

\begin{figure*}[t] 
\includegraphics[width=\textwidth,keepaspectratio]{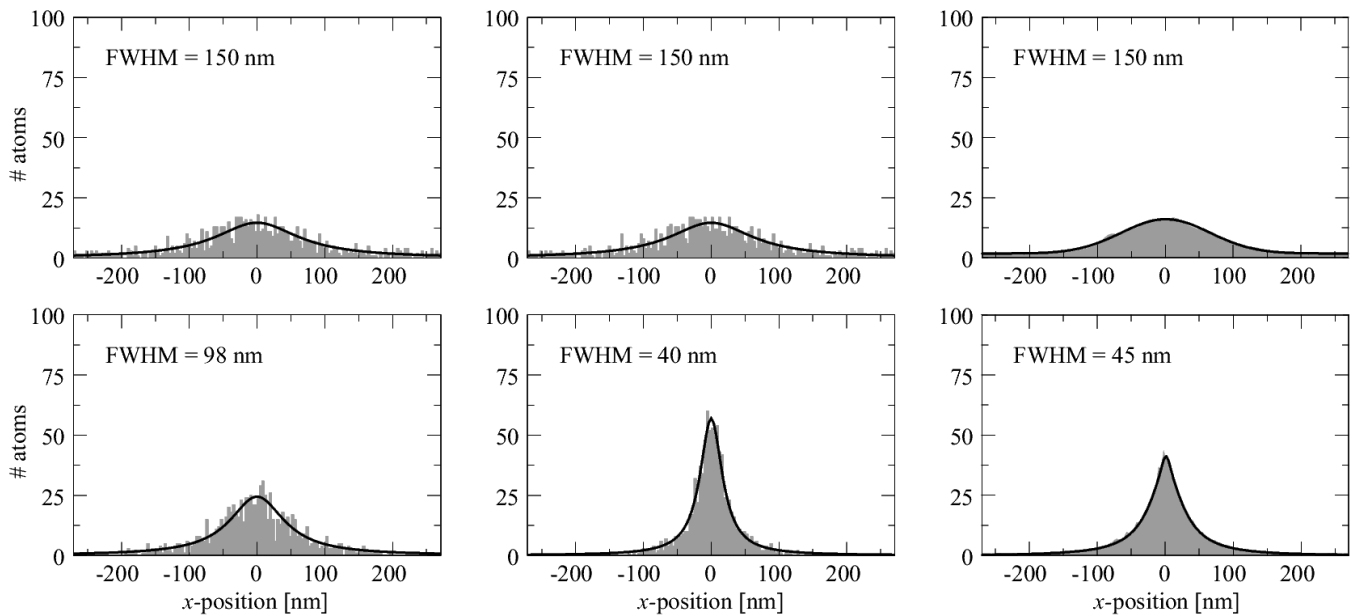} 
\caption {\label{histograms} Histograms of the atom distributions 
taken at $z = w_z/4 = 83~\mu$m. The upper graphs show the 
distributions for the focusing regime in the potential model 
(left), Minogin model (centre), and the MCWF simulations (right). 
Similarly, the channelling regime is represented by the lower 
graphs.} 
\end{figure*} 

Distribution plots of the atoms are shown in Fig.~\ref{xyplots}.
Every dot in these graphs represents a position where the atom has
hit the sample after travelling through the standing-wave light
field. The calculations are performed on atoms in a lattice of
$101 \times 101$ atoms. The starting positions of the atoms range
from $(x,y) = (-\lambda/2,-2w_y)$ to $(x,y) = (\lambda/2,2w_y)$.
The plots only show half of this window in the $x$-direction, from
$x = -\lambda/4$ to $x = \lambda/4$. In the focusing regime, the
light-field does not act as a perfect lens, but shows some
abberations. This imperfection is enhanced by the longitudinal and
transverse velocity spread of the atoms. As a result, the best
position to place the sample is not at the centre of the light
beam $(z = 0)$, but slightly behind it. For the channelling
regime, the sample position is far less critical, but the best
results are at similar positions. Therefore, the comparison of the
two regimes for the different models in Fig.~\ref{xyplots} is
performed at $z = w_z/4 = 83~\mu$m, where optimal results are
expected.

The first two plots show the atom distributions in the focusing
regime for the potential model (first plot) and the Minogin model
(second plot). As expected from the comparison of the trajectories
in this regime, shown in the upper graphs of Fig.~\ref{traject},
the differences between the two models are negligible. The
distributions in the channelling regime with the potential model
(third plot) and the Minogin model (fourth plot) show that the
atoms are confined for a wider range along the $y$-axis as
compared to the focusing regime. Furthermore, for the Minogin
model, the atoms are more localized to the nodes of the standing
wave.

To make a more quantitative comparison of the four plots of
Fig.~\ref{xyplots}, histograms of the atom distributions are
presented in Fig.~\ref{histograms}. Only the atoms located between
$y = -100~\mu$m and $y = 100~\mu$m are taken into account for the
histograms, because atoms outside this region do not contribute to
the desired pattern (in the focusing regime). The black curves
through the histograms are Lorentzian fits from which the Full
Width at Half Maximum (FWHM) of the distribution can be deduced.
The upper graphs, that represent the calculations in the focusing
regime, show that the FWHM of the distribution is 150~nm for both
semi-classical models as well as for the quantum-mechanical model.
For the calculations in the channelling regime (lower graphs), the
distribution of the potential model (left graph) has a FWHM of
98~nm. This width deviates clearly from the distribution of the
Minogin model (centre graph), which has a FWHM of 40~nm. The
smaller width of this distribution can be explained from the
additional cooling force that is included in the Minogin model.
The distributions for the full quantum-mechanical calculations are
shown in the right graphs. The FWHM of the central peak is 45~nm,
somewhat larger than the results from the continued fraction
semi-classical results. This larger width is mostly due to the
fact that the steady-state situation for the atoms is not
completely realized during the interaction time. Achieving a
steady state requires a number of spontaneous emissions per atom.
Due to the large detuning of the light field from atomic
resonance, and due to the fact that the atoms are channelled in
the region of low light intensity, almost one half of the atoms
never undergoes a spontaneous emission. The influence of diffusion
is seen in the wings of the MCWF distributions. However, overall
the MCWF results agree very well with the semi-classical
calculations using the Minogin model, confirming the large
reduction in the FWHM of the distribution due to the cooling
force.

\section{Conclusions}
\label{concl}

The simulations have shown that nanoscale patterns can be created
utilizing the optical dipole force to guide atoms through a
standing-wave light field. For the calculation of the atomic
motion, two models have been applied: a simple one that uses the
conventional dipole force, derived from a potential, and another
one that includes the velocity dependence of this dipole force. In
the `conventional' focusing regime, where a low-power light field is used, the
differences between the potential model and the Minogin model are
negligible. Both models show that in principle nanoscale pattern
sizes of 150~nm can be achieved with the specified experimental
parameters. It should be noted, that these parameters are not
optimized for focusing: using a smaller laser focus and more laser
power, much tighter focusing can be achieved without entering the
channelling regime. However, the sample position is very critical,
and good focusing is only achieved for atoms that pass the light
field close to its centre. The high-power channelling regime, that we
are presently using for nanostructure production with metastable helium atoms,
is more robust. It is therefore better suitable for
creating narrow patterns. In this regime, the Minogin model shows
strong deviations from the potential model, due to the
contribution of velocity dependent terms to the dipole force.
Because of the large transverse velocity $(k v_x > \Gamma)$ of the
atoms in the standing-wave light field, the potential model breaks
down and should not be applied for calculations in this regime.
The results of the calculations with the Minogin model show
structures with a FWHM of 40~nm. This model does not take into
account the momentum diffusion and transient effects, which leads
to broadening of the structure size, as is shown by the MCWF
simulations. However, the MCWF simulations support the results of
the Minogin model by showing a better confinement of the atoms in
the standing-wave light field than expected from calculations with
the conventional model. This is due to an additional cooling force
that is not incorporated in the conventional model.

\section*{Acknowledgements}

Financial support from the Foundation for Fundamental Research on
Matter (FOM) is gratefully acknowledged.

\appendix
\section{Continued fraction method}
\label{calccf}

The determination of the dipole force in the Minogin mod\-el,
given by Eqs.~(\ref{Fcfpar}) and (\ref{Fcfperp}), requires the
optical Bloch vector components $u_n$, that are embedded in the
expressions of the Fourier coefficients in Eqs.~(\ref{parcoef})
and (\ref{perpcoef}). The coefficients $u_n$ can be calculated
with a continued fraction method according to Minogin and Letokhov
\cite{mino87}. A summary of this method is described below.

The upper expression of Eq.~(\ref{recur}) can be rewritten in the form
\begin{equation}
\label{recur_u}
u_n = \frac{\Delta}{\Gamma/2 + i n k v_x} v_n.
\end{equation}
Since $v_n$ is non-zero for odd $n$, and $w_n$ is non-zero for
even $n$, the middle and lower expressions of Eq.~(\ref{recur})
can be combined to a single expression as
\begin{equation}
\label{combicoef}
b_n - D_n ( b_{n-1} + b_{n+1} ) = -\delta_{n0}/2,
\end{equation}
where
\begin{equation}
b_n = \left\{
\begin{array}{l@{\qquad}l}
v_n & \mbox{for odd $n$,} \\
w_n & \mbox{for even $n$,}
\end{array}
\right.
\end{equation}
and the coefficients $D_n$ are given by
\begin{equation}
D_n = \left\{
\begin{array}{l@{\qquad}l}
-\frac{\displaystyle \Omega \left(\Gamma/2 + i n k v_x \right)}
{\displaystyle \left(\Gamma/2 + i n k v_x \right)^2 + \Delta^2} & \mbox{for odd $n$,} \\
\frac{\displaystyle \Omega}{\displaystyle \Gamma + i n k v_x} & \mbox{for even $n$.}
\end{array}
\right.
\end{equation}
The coefficients $b_n$ fulfil the reality condition
\begin{equation}
b_{-n} = b_n^*,
\end{equation}
and therefore only non-negative integers ($n \ge 0$) have to be
considered. The relation between two successive quantities $b_{n}$
and $b_{n+1}$ can then be found by substituting
\begin{equation}
\label{recur_b}
b_{n+1} = q_n b_n
\end{equation}
into Eq.~(\ref{combicoef}). This leads to a recursion relation for
$q_n$ given by
\begin{equation}
q_n = \frac{1}{D_n} - \frac{1}{q_{n-1}}.
\end{equation}
The quantity $q_0$ can now be expressed as a convergent continued fraction
\begin{equation}
q_0 = \frac{D_1}{\displaystyle 1 + \frac{p_1}{\displaystyle 1 +
\frac{p_2}{\displaystyle 1 + \frac{p_3}{\displaystyle 1 +
\ldots}}}},
\end{equation}
where
\begin{equation}
p_n = -D_n D_{n+1}.
\end{equation}
The quantity $b_0 = w_0$ is found by solving Eq.~(\ref{combicoef})
for $n=0$ and gives
\begin{equation}
b_0 = \frac{1}{4\Omega/\Gamma~\mbox{Re}\,(q_0) - 2}.
\end{equation}
From Eq.~(\ref{recur_b}) and this initial coefficient $b_0$, all
successive coefficients $b_n$ can be calculated. By substituting
the coefficients $b_n$ ($= v_n$ for odd $n$) into
Eq.~(\ref{recur_u}), the coefficients $u_n$ can be found
straightforwardly.

\end{document}